\documentclass[twoside,11pt]{article}

\usepackage{graphicx, subfigure}
\usepackage[top=1in,bottom=1in,left=1in,right=1in]{geometry}
\usepackage[sort&compress]{natbib} 
\usepackage{amsfonts, amsmath, amssymb, amsthm, constants, bbm}
\usepackage{MnSymbol}
\usepackage{mathtools}
\usepackage{enumitem}
\usepackage{booktabs}

\usepackage{caption}
\usepackage{color}
\definecolor{darkred}{RGB}{100,0,0}
\definecolor{darkgreen}{RGB}{0,100,0}
\definecolor{darkblue}{RGB}{0,0,150}

\usepackage{hyperref}
\hypersetup{colorlinks=true, linkcolor=darkred, citecolor=darkgreen, urlcolor=darkblue}
\usepackage{url}

\theoremstyle{remark}

\def\beq{\begin{equation}} % \setcounter{equation}{1}}
\def\eeq{\end{equation}}
\def\beqn{\begin{eqnarray*}}
\def\eeqn{\end{eqnarray*}}
\def\Bitem{\begin{itemize}\setlength{\itemsep}{.2in}}
\def\bitem{\begin{itemize}\setlength{\itemsep}{.05in}}
\def\eitem{\end{itemize}}
\def\Benum{\begin{enumerate}\setlength{\itemsep}{.2in}}
\def\benum{\begin{enumerate}\setlength{\itemsep}{.05in}}
\def\eenum{\end{enumerate}}
\def\bmult{\begin{multline*}}
\def\emult{\end{multline*}}
\def\bcenter{\begin{center}}
\def\ecenter{\end{center}}
\def\bframe{\begin{frame}}
\def\eframe{\end{frame}}

\newcommand{\secref}[1]{Section~\ref{sec:#1}}

\newcommand{\tabref}[1]{Table~\ref{tab:#1}}

\newcommand{\conref}[1]{Constraint~\ref{con:#1}}

\DeclareMathOperator*{\argmax}{arg\, max}
\DeclareMathOperator*{\argmin}{arg\, min}
\newcommand{\E}{\operatorname{\mathbb{E}}}

\def\1{\mathbbm{1}}

\definecolor{purple}{rgb}{0.4,.1,.9}

{
  \theoremstyle{plain}
  
}
{
  \theoremstyle{remark}
  
}
{
  \theoremstyle{plain}
  
}
{
  \theoremstyle{plain}
  \newtheorem{constraint}{Constraint}
}
{
  \theoremstyle{plain}
  \newtheorem{proposition}{Proposition}
}

\begin{document}
\thispagestyle{empty}

\title{$K$-Means and Gaussian Mixture Modeling\\ with a Separation Constraint}
\author{
He Jiang\,\footnote{Department of Mathematics and Statistics, California State Polytechnic University Pomona, Pomona, CA,
USA, 91768; hejiang@cpp.edu.} 
\and 
Ery Arias-Castro\,\footnote{Department of Mathematics, University of California San
Diego, La Jolla, CA, USA, 92093; eariascastro@ucsd.edu. } 
}
\maketitle

\begin{abstract}

We consider the problem of clustering with $K$-means and Gaussian mixture models with a constraint on the separation between the centers in the context of real-valued data. We first propose a dynamic programming approach to solving the $K$-means problem with a separation constraint on the centers, building on \citep{wang2011ckmeans}. In the context of fitting a Gaussian mixture model, we then propose an EM algorithm that incorporates such a constraint. A separation constraint can help regularize the output of a clustering algorithm, and we provide both simulated and real data examples to illustrate this point.
\end{abstract}

%\tableofcontents

%%%%%%%% SECTION 1 %%%%%%%%
\section{Introduction} \label{sec:intro}

Cluster analysis is broadly seen as one of the most important tasks in (unsupervised) data analysis \citep{jain1999data, kaufman2009finding}. In some situations, the analyst might have some prior knowledge or relevant information regarding the clusters, and incorporating this information --- which may be done via constraints --- is thought to improve the accuracy of the clustering output \citep{basu2008constrained}. In the present paper, we address situations where the analyst has prior knowledge on the separation between the cluster centers. We indeed focus on methods that rely on centers to define clusters: the $K$-means criterion and Gaussian mixture models (GMM).

\subsection{Constrained $K$-means}
A number of constrained variants of $K$-means have been proposed in the literature, where a significant amount of attention have been placed on the cluster sizes such as imposing a lower bound on the minimum cluster size or aiming for balancing the cluster sizes.
For instance, \cite{bradley2000constrained} addressed the problem of small or even empty clusters by enforcing a size constraint in the cluster assignment step of Lloyd's algorithm; 
\cite{nielsen2014optimal} designed a dynamic $K$-means algorithm in 1D aimed at optimizing the Bregman divergence that can incorporate a constraint on the minimum cluster size; 
and \cite{banerjee2006scalable} proposed an approach to $K$-means that is able to handle cluster size balance constraints.
Constrained $K$-means has also been widely considered in other areas \citep{basu2008constrained}. 
For example, \citep{wagstaff2001constrained, basu2004active} considered forcing some user-specified pairs of observations to belong to the same cluster while forcing others to belong to different clusters. 
%\cite{basu2004active} proposed a new criterion incorporating both within cluster sum of squares and must-link and cannot-link constraints, and then proposed PCKMeans algorithm to minimize that criteria. 
An overview of instance-level constrained clustering, including $K$-means approaches and related methods, can be found at \citep{davidson2007survey}. \cite{davidson2005clustering} also incorporated a minimum separation constraint between clusters mandating that any two points assigned to different clusters need to be separated by at least $\delta$, a parameter specified by the analyst.
\cite{szkaliczki2016clustering} considered points arriving sequentially and constraints where clusters have to be of the form $\{x_i, \dots, x_{i+j}\}$. 
A recent survey of constrained clustering, including $K$-means related methods and modern development can be found at \cite{ganccarski2020constrained}. 

We propose a variant of $K$-means where a user-specified separation between the cluster centers is enforced which, to the best of our knowledge, is novel. We place ourselves in dimension one, where we are able to solve the resulting constrained optimization problem exactly by building on the dynamic programming approach of \cite{wang2011ckmeans}.

%As we were finishing the present paper we found an article where such a separation constraint is considered \citep{usami2014constrained}. The algorithm proposed in that paper is a variant of Lloyd's where clusters are simply merged if the distance between their centers falls below a prescribed threshold. Besides being more ad hoc, in limited experiments we have not found the algorithm to perform well, and thus do not consider it further in our discussion.

\subsection{Constrained Gaussian mixture models}
Model based clustering is an important aspect in clustering \citep{fraley2002model, mclachlan2004finite}, and among parametric finite mixture models, Gaussian mixture models (GMM) are by far the most popular, with their parameters frequently estimated by the EM algorithm or variants \citep{dempster1977maximum, mclachlan2007algorithm}. 

EM-type algorithms forcing various constraints on the parameters of the mixture model have been considered extensively in the literature. 
\cite{kim1995restricted}, in a more general setting that includes GMM, considered enforcing some equality constraints on a subset of the parameters and provided an EM algorithm based on a projected Newton--Raphson step in the maximization stage. 
Along the lines, but addressing both equality and inequality constraints, \cite{jamshidian2004algorithms} used a projected gradient step instead, in conjunction with an active set method in order to handle the inequality constraints. 
See also \citep{takai2012constrained}. 
In the context of linear models in regression, EM-type algorithms were also proposed to handle constraints on the parameters, in particular linear equalities and inequalities \citep{shin2001testing, shi2005restricted, zheng2005restricted, tan2007fast, tian2008type, zheng2012likelihood, davis2012testing}. 

In the context of GMM proper, constraints have been considered for a very long time to reduce the number of parameters, for example, by making all the variances or covariance matrices to be equal.
Beyond that, constraints on the variances or eigenvalues of the covariance matrices have been considered extensively, mostly for the purpose of bounding the log likelihood \citep{day1969estimating}. 
In particular, \cite{hathaway1985constrained} proposed a lower bound on the ratio of any two component standard deviations, which enabled the author to prove the strong consistency of the resulting maximum likelihood estimator. 
This was followed by \citep{hathaway1986constrained} where a constrained EM algorithm was proposed for maximizing the likelihood with this lower bound constraint on the ratio of any two standard deviations, and also another lower bound constraint on the weights. 
Generalizing Hathaway's results to higher dimensions, \cite{ingrassia2004likelihood, ingrassia2007constrained, ingrassia2011degeneracy, rocci2018data} examined EM algorithms enforcing constraints on the eigenvalues of the covariance matrices.  

In addition to constraints on component variance or eigenvalues, estimation of component means under constraints has also been considered in the literature, where EM-type algorithms have been proposed for that purpose. For example, \cite{pan2007penalized} and \cite{wang2008variable} considered penalizing the $L_1$ and $L_{\infty}$ norms of the mean vectors, respectively, to enforce sparsity and/or regularize the model in high dimensions, while \cite{chauveau2013ecm} and \cite{qiao2015gaussian} proposed EM algorithms for linear constraints on the mean parameters.

We consider here, in dimension 1, arbitrary constraints on the separation between adjacent means. We propose an EM algorithm that enforces such constraints, where the M step amounts to solving a quadratic program.

\subsection{Setting and content}
The main focus will be on real-valued data. We do not know how to enforce a separation constraint on the centers in higher dimensions, at least not in such principled fashion.
The data points will be assumed ordered without loss of generality, denoted $x_1 \le \cdots \le x_N$ and gathered in a vector $x = (x_1, \dots, x_N) \in \mathbb{R}^N$. 
Our goal will be to group these data points into $K$ clusters, where $K$ is specified by the user. 

The organization of the paper will be as follows. 
%In \secref{basics}, we define basic concepts and provide an illustrative example where a constraint on the center separation significantly improves the results of both $K$-means and GMM. 
In \secref{K-means}, we introduce our dynamic programming algorithm for exactly solving $K$-means in dimension one with a constraint on the minimum separation between the centers. 
In \secref{GMM}, we introduce our EM algorithm for (approximately) fitting a Gaussian mixture model by maximum likelihood under the same constraint. It is a true EM algorithm in that the likelihood increases with each iteration. 
In \secref{numerics}, we describe some numerical experiments on both simulated data and real data to illustrate the use and accuracy of our methods.

%%%%%%%% SECTION 2 %%%%%%%%
\section{A Dynamic Program for Separation-Constrained $K$-Means}
\label{sec:K-means}

Given data points on the real line, $x_1, \dots, x_n$, and a desired number of cluster, $K$, basic $K$-means is defined as the following optimization problem \citep{macqueen1967some}
\begin{align*}
\text{minimize}& \quad \sum_{i=1}^n \min_{k=1, \dots, K} (x_i - \mu_k)^2, \\
\text{over}& \quad \mu_1 < \cdots < \mu_K.
\end{align*}
This problem is difficult in general. For one thing, a brute-force approach by grid search would be in dimension $K$ and quickly unfeasible or too costly.
The problem is instead most often approached via iterative methods such as Lloyd's algorithm \citep{lloyd1982least}, which consists in alternating between, for all $k$, defining Cluster $k$ as the set of points nearest to $\mu_k$ and then recomputing $\mu_k$ as the barycenter or average of the points forming Cluster $k$.
In dimension 1, however, the $K$-means problem can be solved by dynamic programming as was established and carried out by \cite{wang2011ckmeans}.

We are interested in a variant of the $K$-means problem where the following separation constraint on the centers is enforced:
\begin{constraint}
\label{con:separation}
$\mu_{k+1} - \mu_{k} \geq \delta$ for all $k = 1, \dots, K-1$.
\end{constraint}
In other words, we consider the following optimization problem:
\begin{align*}
\text{minimize}& \quad \sum_{i=1}^n \min_{k=1, \dots, K} (x_i - \mu_k)^2, \\
\text{over}& \quad \mu_1 < \cdots < \mu_K \quad \text{satisfying} \quad \mu_{k+1} - \mu_{k} \ge \delta \ \text{for all } k = 1, \dots, K-1.
\end{align*}
The amount of separation is determined by the user-specified parameter $\delta \ge 0$. (Of course, if $\delta = 0$, we recover the basic $K$-means problem.)
Inspired by the work of \cite{wang2011ckmeans}, we propose a dynamic programming algorithm to solve this constrained variant of $K$-means.
The remaining of this section is dedicated to describing this algorithm, which is encapsulated in \tabref{DP_sep} and \tabref{DP_sep_routine}.

\subsection{The DP algorithm of \cite{wang2011ckmeans}}
Given the sorted data $x = (x_1, \dots, x_N)$, for $r_1=1,\dots,N$ and $r_2=r_1,\dots,N$, we first define the within current cluster center ($C$) and the within current cluster sum of squares ($W$):
\begin{equation}
\label{C}
    C[r_1,r_2] = \frac{1}{r_2-r_1+1} \sum_{i=r_1}^{r_2} x_i,
\end{equation}

\begin{equation}
\label{W}
    W[r_1,r_2] = \frac{1}{r_2-r_1+1} \sum_{i=r_1}^{r_2} (x_i - C[r_1,r_2] )^2 
\end{equation}
Each element can be efficiently computed using a simple recursion. For $i = 1, \dots, N$ and $m = 1, \dots, K$, define $D[i,m]$ as recording the minimum error sum of squares when grouping the first $i$ data points into $m$ clusters. The main idea of \cite{wang2011ckmeans} is to update $D$ dynamically as follows:
\begin{equation}
\label{DP}
    D[i,m] = \min_{j=m,\dots,i}\ D[j-1,m-1] + W[j,i] \ .
\end{equation}
This is carried out for $i=1,\dots,N$ and $m=1,\dots,K$.
(Initialization is $D[0,m] = 0$ for all $m$ and $D[i,0] = 0$ for all $i$.)

The method of \citep{wang2011ckmeans} was designed for solving the basic $K$-means problem and needs to be modified, in a substantial way, to solve the separation-constrained $K$-means problem.
For illustrative purposes, consider clustering data $x = (-2,1,2,4,5,6,9,10)$ into $K=5$ clusters, while keeping each cluster center separated by at least $\delta = 1.75$. Note that without separation the clustering assignment is $ 1 2 2 3 3 4 5 5$ but the centers of Cluster $3$ and Cluster $4$ are only separated by $1.5$, which is less than $\delta$. With separation, the clustering assignment is $1 2 3 3 4 4 5 5$.
The same example is used to highlight two issues that need to be addressed in order to incorporate the separation constraints:

\begin{itemize}
    \item The $m$-th cluster consisting of $x_j, \dots, x_i$, and the previous cluster, i.e., the $(m-1)$-th cluster that gives the optimal clustering of $j-1$ data points into $m-1$ clusters, might not be separated by $\delta$ in terms of their means.
    In our example, when grouping $\{x_1,\dots, x_5\}$ into $3$ clusters, the optimal solution is $12233$, but when we try to group $\{x_1,\dots, x_6\}$ into $4$ clusters under the separation constraint it becomes impossible to do so as $C[6,6] - C[4,5] < \delta$. (We are not allowing the clusters to be empty.)

    \item Consider the stage where we have grouped $x_1, \dots, x_i$ into $m$ clusters, and the last cluster is $x_j, \dots, x_i$. Then it is not necessarily the case that the first $m-1$ clusters constitute an optimal grouping of $x_1, \dots, x_{j-1}$, and so the recursion \eqref{DP} is not necessarily valid. 
    In our example, the optimal clustering for grouping $\{x_1,\dots,x_4\}$ into $3$ clusters is $1223$, however when grouping $\{x_1,\dots,x_5\}$ into $4$ clusters under the separation constraint the clustering assignment becomes $12334$, so that the starting point of cluster $3$ has changed from $x_4$ to $x_3$.
\end{itemize}

In brief, \eqref{DP} cannot be used as the updating rule. 
Although the first issue is easy fix by checking that the separation constraint is satisfied, the second issue is more difficult to deal with and requires us to introduce additional matrices/tensors as described below.

\subsection{Our DP algorithm with separation constraint}
We describe here our algorithm.
First initialize:
\begin{equation}
\label{DP_sep_intial1}
  \beta = W[1,N] + 1, \qquad D = \beta\, 1_{N \times K}, \qquad I = 0_{N \times K}, \qquad B = 0_{N \times K}, \qquad U = \beta\, 1_{N \times N \times K}\ ,
\end{equation}
where $\beta$ is a placeholder value chosen to be larger than the maximum possible error sum of squares.
Define:
\begin{equation}
\label{DP_sep_intial2}
\begin{gathered}
  D[i,1] = \sum_{l=1}^i (x_l - C[1,i])^2, \quad I[i,1] = 1, \quad B[i,1] = 1, \quad U[i,r,1] = 0, \\ \text{for } i=1,2,\dots,N;\ r=i,\dots,N\ .
\end{gathered}
\end{equation}
The $D$ matrix is used to store the optimal error sum of squares under the separation constraint, specifically $D[i,m]$ records the minimal error sum of squares when putting the first $i$ values into $m$ clusters while satisfying \conref{separation}, if possible (recall we do not allow clusters to be empty).  
The corresponding entry in the $I$ matrix, $I[i,m]$, is used to record the index of the first data point in the $m$-th cluster in the optimal clustering of the first $i$ data points into $m$ clusters (the clustering corresponding to $D[i,m]$).
$B[i,m]$, on the other hand, is used to store the smallest possible index of the leftmost data point in the $m$-th cluster over all groupings of the first $i$ data points into $m$ clusters satisfying the separation constraint. 
Lastly, $U[q,r,m]$ records the optimal error sum of squares for grouping the first $q-1$ data points into $m-1$ clusters, while satisfying the constraint that the $(m-1)$-th cluster and $m$-th cluster, which consists of data points $x_q, \dots, x_r$, are separated by $\delta$ in means. 

In our updates of the above matrices/tensors, we enforce the separation constraint. 
Suppose that $D[\cdot,m-1], I[\cdot,m-1], B[\cdot,m-1]$ as well as $U[\cdot,\cdot,m-1]$ have been updated, so we are at stage $m$ with the immediate task of updating $D[\cdot,m], I[\cdot,m], B[\cdot,m]$ as well as $U[\cdot,\cdot,m]$.
Now consider grouping the first $i$ data points into $m$ clusters. 
% If this is not possible without violating the separation constraint, then $D[i,m]=\beta$, $I[i,m]=B[i,m]=0$, and $U[i+1,r,m+1]=\beta$.
If $B[i-1,m-1] = 0$, do nothing (the values stay at the initial ones). 
Otherwise, for each $j = m, \dots, i$, we find the first value in the descending sequence $T_j \in \{I[j-1,m-1],\dots,B[j-1,m-1]\}$ that satisfies $C[j,i] - C[T_j,j-1] \geq \delta$. Note that $T_j$ also depends on $i$ and $m$, but that is left implicit. The reason for defining $T_j$ as such is because starting from $B[j-1,m-1]$, the closer we are to $I[j-1,m-1]$, the smaller the error sum of squares. Then the main update that guarantees optimality is:

\begin{equation}
\label{DP_sep}
    D[i,m] = \min_{j = m, \dots, i}\ \bigg{(} U[T_j,j-1,m-1] + W[T_j,j-1] + W[j,i] \bigg{)} \ ,
\end{equation}
and 
\begin{equation}
    U[j,i,m] = U[T_j,j-1,m-1] + W[T_j, j-1]\ .
\end{equation}
Then $I[i,m]$ is updated as the smallest minimizing index in \eqref{DP_sep} and $B[i,m]$ is updated as the smallest $j$ that satisfies $C[j,i] - C[T_j,j-1] \ge \delta$.

\paragraph{Computational complexity}
While the algorithm of \cite{wang2011ckmeans} has time complexity $O(N^2 K)$, our method has computational complexity $O(N^3 K)$. While this is true in the worst case, in practice the computational cost appears much lower as moving from $I[j-1, m-1]$ to $B[j-1, m-1]$ and stopping at the first value where the constraint is satisfied considers significantly fewer options than $N$. The upper bound for space complexity of this algorithm is $O(N^2 K)$ --- compared to $O(N K)$ for the algorithm of \cite{wang2011ckmeans}. This is due to the maintaining of the tensor $U$, but again, this is in the worst case, as in practice $U$ is very sparse (most of its coefficients are equal to the initial value $\beta$) and the space complexity is far below the upper bound. 

We are now ready to use \tabref{DP_sep_routine} to find the the optimal error sum of squares under \conref{separation} and find the corresponding optimal clustering using \tabref{DP_sep}. We note that the result $b$ of \tabref{DP_sep} is a vector indicating the index of each cluster's last datapoint. We also note that in \tabref{DP_sep_routine} there is the possibility that the minimum value of $p$ occurs at multiple locations; when this happens, we select the entry with the smallest location index.

% \begin{algorithm}[!th]
% \caption{\quad \bf Basic rejection sampling}
% \label{alg:rejection-sampling}
% \begin{algorithmic}
% \STATE \textbf{Input:} target density $f$, proposal density $f_0$, constant $c$ satisfying \eqref{rejection-bound}. 
% \STATE \textbf{Output:} one realization from $f$
% \medskip
% \STATE \textbf{Repeat:} generate $y$ from $f_0$ and $u$ from $\Unif([0,1])$, independently
% \STATE \textbf{Until} $u \le f(y)/c f_0(y)$
% \STATE \textbf{Return} the last $y$ 
% \end{algorithmic}
% \end{algorithm}

%%%%%%%%%%%%%%%%%%%%%%%%
%%%%%%%%%%%%%%%%%%%%%%%%
\begin{table}
\centering
\caption{Center separation constrained $K$-means}
\label{tab:DP_sep}
\bigskip
\setlength{\tabcolsep}{0.22in}
\begin{tabular}{ p{0.9\textwidth}  }
\toprule
%\multicolumn{2}{c}{Item} \\
%\cmidrule(r){1-2}
{\textbf{inputs}: $C, I, B$ (all three obtained from \tabref{DP_sep_routine}), number of clusters $K$, separation $\delta$} \\ \midrule

initialize $b = 0_{\{K\}}$, compute $b[K] = I[N,K]$, and set $l = b[K] - 1$

\textbf{if} $K=1$ \textbf{then}

\hspace{3mm} \textbf{return} $b$

\textbf{for} $j=I[l,K-1],\dots,B[l,K-1]$ \textbf{do}

\hspace{3mm} \textbf{if} $C[ l+1, N ] - C[j,l] \geq \delta$ \textbf{then}
   
\hspace{3mm} \hspace{3mm} $b[K-1] = j, l = j-1$

\hspace{3mm} \hspace{3mm} \textbf{break}

\textbf{if} $K=2$ \textbf{then}

\hspace{3mm} \textbf{return} $b$

\textbf{for} $k=(K-2),\dots,1$ \textbf{do}

\hspace{3mm} \textbf{for} $j = I[l,k], \dots , B[l,k]$ \textbf{do}

\hspace{3mm} \hspace{3mm} \textbf{if} $(C[ l+1, b[k+2]-1 ] - C[j,l]) \geq \delta$ \textbf{then}

\hspace{3mm} \hspace{3mm} \hspace{3mm} $b[k] = j, l = j-1$

\hspace{3mm} \hspace{3mm} \hspace{3mm} \textbf{break}
       
\textbf{return} $b$ \\

\bottomrule
\end{tabular}
\end{table}

%%%%%%%%%%%%%%%%%%%%%%%%
%%%%%%%%%%%%%%%%%%%%%%%%

%%%%%%%%%%%%%%%%%%%%%%%%
%%%%%%%%%%%%%%%%%%%%%%%%
\begin{table}
\centering
\caption{Routine computing  matrices $C, W, D, I, B$}
\label{tab:DP_sep_routine}
\bigskip
\setlength{\tabcolsep}{0.22in}
\begin{tabular}{ p{0.9\textwidth}  }
\toprule
%\multicolumn{2}{c}{Item} \\
%\cmidrule(r){1-2}
{\textbf{inputs}: data $x_1 \le \cdots \le x_N$, number of clusters $K$, separation $\delta$} \\ \midrule

compute $C$ and $W$ as in \eqref{C} and \eqref{W} \\
define $\beta$ and initialize $D, I, B, U$ as in \eqref{DP_sep_intial1} and \eqref{DP_sep_intial2}

%% K=1 case
\textbf{if} $K=1$ \textbf{then}

go to return step
\hspace{3mm} 

%% K=2 case
\textbf{else if} $K=2$ \textbf{then}

%\hspace{3mm} do everything in $K=1$ case except go to return step, and:

\hspace{3mm} \textbf{for} $i=2,\dots,N$ \textbf{do}

\hspace{3mm} \hspace{3mm} $ p = \beta_{\{i - 1\}}  $

\hspace{3mm} \hspace{3mm} \textbf{for} $j=2,\dots,i$ \textbf{do} 

\hspace{3mm} \hspace{3mm} \hspace{3mm} \textbf{if} $C[j,i] - C[1,j-1] \geq \delta$ \textbf{then}

\hspace{3mm} \hspace{3mm} \hspace{3mm} \hspace{3mm} $p[j - 1] = W[j,i] + W[1,j-1] $

\hspace{3mm} \hspace{3mm} \hspace{3mm} \hspace{3mm} $U[j,i,2] = W[1,j-1]$
      
\hspace{3mm} \hspace{3mm}  $D[i, 2] = \min(p)$

\hspace{3mm} \hspace{3mm} \textbf{if} $D[i, 2] = \beta$ \textbf{then}

\hspace{3mm} \hspace{3mm} \hspace{3mm} $I[i, 2] = 0, B[i, 2] = 0$

\hspace{3mm} \hspace{3mm} \textbf{else} 

\hspace{3mm} \hspace{3mm} \hspace{3mm} $I[i, 2] = \min(\argmin(p))+1$, $B[i, 2] = \min\{a: p[a] < \beta\}+1$

\hspace{3mm} \hspace{3mm} go to return step

%% K>=3 general case
\textbf{else} 

\hspace{3mm} do everything in $K=2$ case except go to return step, and:

\hspace{3mm} \textbf{for} $m=3,\dots,K$ \textbf{do}

\hspace{3mm} \hspace{3mm} \textbf{for} $i=m,\dots,N$ \textbf{do}

\hspace{3mm} \hspace{3mm} \hspace{3mm} \textbf{if} $D[i-1, m-1] < \beta$ \textbf{then}

\hspace{3mm} \hspace{3mm} \hspace{3mm} \hspace{3mm} $p = \beta_{\{i - m + 1\}} $

\hspace{3mm} \hspace{3mm} \hspace{3mm} \hspace{3mm} \textbf{for} $j=m,\dots,i$ \textbf{do}

\hspace{3mm} \hspace{3mm} \hspace{3mm} \hspace{3mm} \hspace{3mm} \textbf{if} $B[j-1, m-1] \neq 0$ \textbf{then} 

\hspace{3mm} \hspace{3mm} \hspace{3mm} \hspace{3mm} \hspace{3mm} \hspace{3mm} \textbf{if} $C[j,i] - C[ I[j-1,m-1] , j-1 ] \geq \delta$ \textbf{then}

\hspace{3mm} \hspace{3mm} \hspace{3mm} \hspace{3mm} \hspace{3mm} \hspace{3mm} \hspace{3mm} $p[j - m + 1] = D[j-1, m-1] + W[j,i]$

\hspace{3mm} \hspace{3mm} \hspace{3mm} \hspace{3mm} \hspace{3mm} \hspace{3mm} \hspace{3mm} $U[j,i,m] = D[j-1, m-1]$

\hspace{3mm} \hspace{3mm} \hspace{3mm} \hspace{3mm} \hspace{3mm} \hspace{3mm} \textbf{else}

\hspace{3mm} \hspace{3mm} \hspace{3mm} \hspace{3mm} \hspace{3mm} \hspace{3mm} \hspace{3mm} \textbf{for} $t = I[j-1,m-1],\dots,B[j-1,m-1]$ \textbf{do}

\hspace{3mm} \hspace{3mm} \hspace{3mm} \hspace{3mm} \hspace{3mm} \hspace{3mm} \hspace{3mm} \hspace{3mm} \textbf{if} $C[j,i] - C[t,j-1] \geq \delta$ \textbf{then} 

\hspace{3mm} \hspace{3mm} \hspace{3mm} \hspace{3mm} \hspace{3mm} \hspace{3mm} \hspace{3mm} \hspace{3mm} \hspace{3mm} \text{ } $p [j - m + 1] = \min(p[j - m + 1], U[t,j-1,m-1] + W[t,j-1] + W[j,i] )$

\hspace{3mm} \hspace{3mm} \hspace{3mm} \hspace{3mm} \hspace{3mm} \hspace{3mm} \hspace{3mm} \hspace{3mm} \hspace{3mm} $U[j,i,m] = U[t,j-1,m-1] + W[t, j-1]$

\hspace{3mm} \hspace{3mm} \hspace{3mm} \hspace{3mm} \hspace{3mm} \hspace{3mm} \hspace{3mm} \hspace{3mm} \hspace{3mm} \textbf{break}

\hspace{3mm} \hspace{3mm} \hspace{3mm} \hspace{3mm}  $D[i, m] = \min(p)$

\hspace{3mm} \hspace{3mm} \hspace{3mm} \textbf{if} $D[i, m] = \beta$ \textbf{then}

\hspace{3mm} \hspace{3mm} \hspace{3mm} \hspace{3mm} $I[i, m] = 0, B[i, m] = 0$

\hspace{3mm} \hspace{3mm} \hspace{3mm} \textbf{else} 

\hspace{3mm} \hspace{3mm} \hspace{3mm} \hspace{3mm} $I[i, m] = \min(\argmin(p))+m-1$, $B[i, m] = \min\{a: p[a] < \beta\}+m-1$ 

\textbf{return} $D, I, B, C, W, \beta$, U \\

\bottomrule
\end{tabular}
\end{table}

%%%%%%%%%%%%%%%%%%%%%%%%
%%%%%%%%%%%%%%%%%%%%%%%%

%%%%%%%% SECTION 3 %%%%%%%%
\section{An EM Algorithm for Separation-Constrained GMM}
\label{sec:GMM}

A Gaussian mixture model is of the form:

\begin{equation}
\label{GMM}
    \sum_{k=1}^{K} \pi_k f(x; \mu_k, v_k),
\end{equation}
where $f(\cdot; \mu, v)$ is the density of the normal distribution with mean $\mu$ and variance $v$, meaning
\begin{equation}
    f(x; \mu, v) = \frac{1}{\sqrt{2 \pi v}} \exp\left(- \frac{(x - \mu)^2}{2 v}\right),
\end{equation}
and $\pi_1, \dots, \pi_K$ are the mixture weights satisfying
\begin{equation}
\min_k \pi_k > 0, \qquad \sum_k \pi_k = 1.
\end{equation}
For convenience, define $\theta = (\pi, \mu, v) $ where $\pi = (\pi_1, \dots, \pi_K) \in \mathbb{R}^K $, $\mu = (\mu_1, \dots, \mu_K) \in \mathbb{R}^K$, and $v = (v_1, \dots, v_K) \in \mathbb{R}^K$. We then let $g_\theta$ denote the mixture \eqref{GMM}.
The working assumption is that the data points, $x_1, \dots, x_N$, are the realization of a sample of some size $N$ drawn iid from a mixture distribution of the form \eqref{GMM}, where a draw from that distribution amounts to drawing a cluster index in $\{1, \dots, K\}$ according to the probability distribution $\pi$ and then generating a real valued observation according the corresponding normal density --- $f(\cdot; \mu_k, v_k)$ if $k$ is the drawn index value. Given the mixture model \eqref{GMM}, a point $x$ is assigned to the most likely cluster, meaning, to 
\begin{equation}
    \argmax_k\, \pi_k f(x; \mu_k, v_k).
\end{equation}

Thus clustering using a Gaussian mixture model boils down to estimating the parameters of the model. Maximum likelihood estimation\footnote{~ Interestingly, maximum likelihood is a competitive method even though the likelihood is not bounded. 
To make sense of what the EM algorithm does, we believe that a useful perspective is to see it as attempting to maximize the likelihood starting from a reasonable initialization, understanding that the E and M steps prevent the algorithm from arriving at pathological values of the parameters.} is a natural candidate, but difficult to implement, even in dimension 1, as maximizing the likelihood is not a convex problem and a grid search is too costly given the number of parameters ($3K-1$).
The Expectation-Maximization (EM) algorithm of \cite{dempster1977maximum} --- arguably the most famous approach to fitting a Gaussian mixture model --- is an iterative method that attempts to maximize the (log) likelihood:
\begin{equation}
    L(\theta, x) = \sum_{i=1}^N \log g_{\theta} (x_i).
\end{equation}
The main idea is based on introducing cluster assignment variables, $z_1, \dots, z_n$, where $z_i = k$ if $x_i$ was drawn from the $k$-th component of the mixture. These assignment variables are not observed (they are said to be latent), but are useful nonetheless for thinking about the problem.
To that end, let $g_\theta(x, z)$ denote the joint density of $(X_i, Z_i)$. Note that $Z_i$ has distribution $\pi$ and $X_i \mid Z_i = k$ has the normal distribution with mean $\mu_k$ and variance $v_k$.
The algorithm starts with a value of the parameter vector, $\theta^0$, and then alternates between an E step and an M step, until some convergence criterion is satisfied. 
Suppose we are at the $s$-th iterate.
The E step consists in computing the expectation of the log-likelihood for an arbitrary value of $\theta$ with respect to the mixture distribution given by the current parameter estimates:
\begin{equation}
    Q(\theta, \theta^s) := \sum_{i=1}^N \E_{\theta^s}\big[\log g_\theta(X_i, Z_i) \mid X_i = x_i\big].
\end{equation}
It turns out that 
\begin{equation}
\label{E_Q}
    Q(\theta, \theta^s) = \sum_{i=1}^N \sum_{k=1}^K w_{i,k}^{s+1} \log (\pi_k f(x_i; \mu_k, v_k)),
\end{equation}
where
\begin{equation}
\label{E_w}
    w_{i,k}^{s+1} := \frac{\pi_k^s f(x_i; \mu_k^s, v_k^s)}{\sum_{l=1}^K \pi_l^s f(x_i; \mu_l^s, v_l^s)}.
\end{equation}
The M step consists of maximizing the resulting function with respect to $\theta$ to yield the updated values of the parameters:
\begin{equation}
    \theta^{s+1} = \argmax_\theta Q(\theta, \theta^s).
\end{equation}
This amounts to the following:
\begin{equation}
\label{M_pi}
\pi_k^{s+1} = \dfrac{1}{N} \sum_{i=1}^N w_{i,k}^{s+1}
\end{equation}
\begin{equation}
\label{M_mu}
\mu_k^{s+1} = \dfrac{1}{\sum_{i=1}^N w_{i,k}^{s+1}} \sum_{i=1}^N w_{i,k}^{s+1} x_i 
\end{equation}
\begin{equation}
\label{M_v}
{v_k}^{s+1} = \dfrac{1}{\sum_{i=1}^N w_{i,k}^{s+1}}  \sum_{i=1}^N w_{i,k}^{s+1} (x_i - \mu_k^{s+1})^2
\end{equation}

% \begin{remark}
% Interestingly, maximum likelihood is a competitive method even though the likelihood is not bounded. 
% %As discussed in \cite[Example 5.51]{van2000asymptotic}, maximizing the likelihood near a ``good estimate" --- often done by taking a likelihood gradient step --- can still be justified. 
% To make sense of what the EM algorithm does, we believe that a useful perspective is to see it as attempting to maximize the likelihood starting from a reasonable initialization, understanding that the E and M steps prevent the algorithm from arriving at pathological values of the parameters. 
% \end{remark}

We are interested in a variant of the EM algorithm for maximizing the likelihood under the following separation constraint on the centers:
\begin{constraint}
\label{con:separation2}
$\delta_{k,1} \leq \mu_{k+1} - \mu_k \leq \delta_{k,2}$
for all $k = 1,\dots,K-1$.
\end{constraint}
The separation parameters, $\delta_{k,1}$ and $\delta_{k,2}$, are set by the user.
We note that this reduces to \conref{separation} when we set $\delta_{1,1} = \dots = \delta_{K-1,1} = \delta$ and  $\delta_{1,2} = \dots = \delta_{K-1,2} = \infty$ (in practice, a very large number). 
We propose an EM algorithm that incorporates \conref{separation2}. It is a true EM algorithm in that the likelihood increases with each iteration. The remaining of this section is devoted to introducing the algorithm, which is otherwise compactly given in \tabref{EM}.

In the algorithm, the E Step is identical to that in the regular EM algorithm. The M Step is also the same except that the maximization is done under \conref{separation2}. 
The constraint is only on the centers, and to isolate that, we adopt an ECM approach \citep{meng1993maximum}, where we first maximize over the weights, which amounts to the same update \eqref{M_pi}; then maximize over the centers subject to \conref{separation2} (see details below); and finally maximize over the variances, which amounts to the same update \eqref{M_v}.

\begin{proposition}
The separation-constrained EM algorithm does not decrease the log likelihood of the model, i.e., $L(\theta^{s+1}, x) \geq L(\theta^{s}, x)$ for all $s$.
\end{proposition}
A proof of this result, although standard, is given in the appendix.

\medskip
The maximization over the centers amounts to solving the following optimization problem:
\begin{equation}
    \label{M_mu_sep}
\begin{split}
\text{minimize} \quad & 
\sum_{i=1}^{N} \sum_{k=1}^{K} \frac{w_{i,k}^{s+1}}{2 v_k^s} ( x_i - \mu_k)^2
\\
\text{over} \quad & \mu_1, \dots, \mu_K \quad \text{such that} \quad 
\delta_{k,1} \leq \mu_{k+1} - \mu_{k} \leq \delta_{k,2} \text{ for all } k = 1,2,\dots,K-1.
\end{split}
\end{equation}
Define $g_{i,k} = w_{i,k}^{s+1}/2 v_k^{s}$ and
\begin{equation}
    G_i = {\rm diag}(g_{i,1}, \dots, g_{i,K}), \qquad G = \sum_{i=1}^{n} G_i\ .  
\end{equation}
Also define
\begin{equation}
    A = \begin{pmatrix} -1 & 1 & 0 & \dots & 0 & 0\\ 0 & -1 & 1 & \dots &0 & 0 \\ \vdots&\vdots&\vdots&\vdots& \vdots& \vdots \\ 0 & 0 & 0 &\dots &-1 & 1 \\ 1 & -1 & 0 & \dots & 0 & 0\\ 0 & 1 & -1 & \dots &0 & 0 \\ \vdots&\vdots&\vdots&\vdots& \vdots& \vdots \\ 0 & 0 & 0 &\dots &1 & -1 \end{pmatrix}, \qquad b_0 = \begin{bmatrix} \delta_{1,1} \\ \delta_{2,1} \\ \dots \\ \delta_{K-1,1} \\ -\delta_{1,2} \\ -\delta_{2,2} \\ \dots \\ -\delta_{K-1,2}  \end{bmatrix}\ ,
\end{equation}
where $A$ is a $2K-2$ by $K$ matrix and $b_0$ is a $2K-2$ vector. With this notation, and let $1^\top$ denote a vector of length $K$ with each entry as $1$, the optimization problem \eqref{M_mu_sep} can be equivalently expressed as follows:
\begin{equation}
\label{Interior_Point}
\begin{aligned}
\text{minimize} \quad & 
\mu^\top G \mu - 2 \sum_{i=1}^{N} x_i\, 1^\top G_i \mu
\\
\text{subject to} \quad & 
    A \mu \geq b_0,
\end{aligned}
\end{equation}
which is a quadratic program.
In practice, we solve it using the primal dual interior point method of \cite{goldfarb1982dual, goldfarb1983numerically} implemented in the \textsf{quadprog} package\footnote{~\url{https://cran.r-project.org/web/packages/quadprog/}} in \textsf{R}.

%%%%%%%%%%%%%%%%%%%%%%%%
%%%%%%%%%%%%%%%%%%%%%%%%
\begin{table}
\centering
\caption{Separation Constrained EM Algorithm for GMM}
\bigskip
\setlength{\tabcolsep}{0.22in}
\begin{tabular}{ p{0.85\textwidth}  }
\toprule
%\multicolumn{2}{c}{Item} \\
%\cmidrule(r){1-2}
{\textbf{inputs}: data $x_1, \dots, x_N$, number of clusters $K$, tolerance value $\gamma$} \\ 

\midrule

\textbf{initialization} \\
initialize $(\pi^1, \mu^1, v^1)$ using constrained 1D optimal $K$-means algorithm as in Section \ref{sec:K-means}, if given the same lower bound constraint; otherwise initialize using 1D optimal $K$-means of \citep{wang2011ckmeans}:
\begin{equation}
    \mu_1^1 < \mu_2^1 <\dots< \mu_K^1, \quad \mu_k^1 =  \frac1{|J_k|} \sum_{i \in J_k} x_i, \quad \pi_k^1 = \frac{|J_k|}N, \quad v_k^1 = \frac1{|J_k|} \sum_{i \in J_k} (x_i - \mu_k)^2
\end{equation}
\noindent where $J_k$ denotes the indices that are clustered into the $k-th$ cluster

\medskip
\textbf{E step} \\
compute $w_{i,k}$ for $i = 1,2,\dots,N$ and $k=1,2,\dots K$:
\begin{equation}
w_{i,k}^{s+1} = \dfrac{\pi_k^{s} N(x_i | \mu_k^s, v_k^s) }{\sum_{j=1}^K \pi_j^{s} N(x_i | \mu_j^s, v_j^s) }
\end{equation}

\textbf{M step} \\
update $\pi_k$, $k=1,2,\dots,K$:
\begin{equation}
\pi_k^{s+1} = \dfrac{1}{N} \sum_{i=1}^N w_{i,k}^{s+1} 
\end{equation}

\noindent update $\mu_k$, $k=1,2,\dots,K$ by the solution to minimization problem \eqref{Interior_Point} \\

\noindent update $v_k$, $k=1,2,\dots,K$:
\begin{equation}
{v_k}^{s+1} = \dfrac{1}{\sum_{i=1}^N w_{i,k}^{s+1}} \sum_{i=1}^N w_{i,k}^{s+1} (x_i - \mu_k^{s+1})^2
\end{equation} \\

\textbf{convergence check}

\textbf{if} $\max\{\max_k {|\pi_k^{s+1} - \pi_k^{s}|},\, \max_k {|\mu_k^{s+1} - \mu_k^{s}|},\, \max_k|v_k^{s+1} - v_k^{s}| \} \le \gamma$

\hspace{3mm} \textbf{then} stop and return $\theta^{s+1} = (\pi^{s+1}, \mu^{s+1}, v^{s+1})$

\noindent \textbf{else}

\hspace{3mm} $s=s+1$ and go to E step \\

\bottomrule
\end{tabular}

\label{tab:EM}

\end{table}
%%%%%%%%%%%%%%%%%%%%%%%%
%%%%%%%%%%%%%%%%%%%%%%%%

%%%%%%%% SECTION 4 %%%%%%%%
\section{Experiments}
\label{sec:numerics}

This section will provide some simulated and real data examples illustrating the application of our algorithms. The constrained algorithms are shown to outperform their unconstrained counterparts in both parameter estimation and in producing a clustering that is more similar to the actual clusters, of course, if the separation constraint reflects the reality of the situation.
Our code is publicly available.\footnote{~\url{https://github.com/h8jiang/Center-Separation-Constrained-K-Means-and-EM-Algorithm/}}
\subsection{Simulations}

We start with experiments performed on simulated data using our constrained algorithms. 
We will first look at the performance of our constrained $K$-means algorithm over regular $K$-means as implemented by \cite{wang2011ckmeans}. 
We then look at the performance of our constrained EM algorithm for fitting Gaussian mixture models with the regular EM algorithm \citep{dempster1977maximum}.
Finally, we examine the impact of the imposed constraints, including when they are not congruent with the actual situation. 

All the simulated data sets were generated from one of the following models:

\begin{itemize}
    \item Model A: $0.333 \text{ }\mathcal{N} (0, 1^2) + 0.667 \text{ }\mathcal{N}(2, 1^2) $  
    \item Model B: $0.45 \text{ }\mathcal{N} (0, 0.75^2) + 0.1 \text{ }\mathcal{N}(2, 1.5^2) +  0.45 \text{ }\mathcal{N}(4, 0.75^2) $ 
    \item Model C: $0.2 \text{ }\mathcal{N} (0, 1^2) + 0.2 \text{ }\mathcal{N}(2, 1^2) +  0.2 \text{ }\mathcal{N}(4, 1^2) + 0.2 \text{ }\mathcal{N}(6, 1^2) + 0.2 \text{ }\mathcal{N}(8, 1^2)$ 
    \item Model D: $0.1 \text{ }\mathcal{N} (0, 0.25^2) + 0.2 \text{ }\mathcal{N}(2, 0.75^2) +  0.4 \text{ }\mathcal{N}(4, 1.25^2) + 0.2 \text{ }\mathcal{N}(6, 0.75^2) + 0.1 \text{ }\mathcal{N}(8, 0.25^2)$  
\end{itemize}

We first apply constrained and unconstrained $K$-means on data generated from Model D (Experiment 1) and on data generated from Model B (Experiment 2). The comparison criteria are total error on cluster center values and the total difference on the cluster sizes:
\begin{equation}
    \sum_{k=1}^K |\hat{\mu_k} - \mu_k|, \quad \text{and} \quad \sum_{k=1}^K |\hat{n_k} - n_k|\ ,
\end{equation}
and also the Rand index. 
The sample size was set to $N=500$, the separation to $\delta=1.95$, and the number of repeats to $R=1000$. 
Note that the separation is satisfied in actuality, and this is important.
We record the mean value over all $R$ experiments and their standard deviations in \tabref{Exp_1_2}. 
It is quite clear that in these experiments constrained $K$-means improves significantly over unconstrained $K$-means in terms of both center estimates and clustering accuracy.

\begin{table}
\centering\small
\caption{Errors and Rand Indices of Optimal $K$-means and Optimal Constrained $K$-means}
\label{tab:Exp_1_2}
\bigskip
\setlength{\tabcolsep}{0.2in}
\begin{tabular}{p{0.15\textwidth} p{0.1\textwidth} p{0.12\textwidth} p{0.12\textwidth} p{0.12\textwidth} }
\toprule
%\multicolumn{2}{c}{Item} \\
%\cmidrule(r){1-2}
{\bf Experiment} & {\bf Model} & {\bf Criteria} & {\bf $K$-means} & {\bf Constrained $K$-means} \\ \midrule
Experiment 1 & Model D & center error & 1.092 (0.276) & 0.374 (0.161) \\ %\midrule
\text{   } & \text{   } & size error & 165.6 (22.9) & 119.9 (25.4)\\ %\midrule
\text{   } & \text{   } & Rand index & 0.786 (0.015) & 0.807 (0.014)\\ \midrule
Experiment 2 & Model B & center error & 1.339 (0.393) & 0.561 (0.190) \\ %\midrule
\text{   } & \text{   } & size error & 143.4 (45.9) & 58.1 (18.8)\\ %\midrule
\text{   } & \text{   } & Rand index & 0.834 (0.016) & 0.858 (0.014)\\

%Experiment 1 & size error & 165.644 (22.941) & 119.898 (25.439)\\ \midrule

\bottomrule
\end{tabular}
\end{table}

Experiments 3, 4 and 5 will show three examples where applying our constrained EM algorithm with a valid separation constraint results in a large improvement of parameter estimation over the regular EM algorithm. Both algorithms are identically initialized using the constrained $K$-means algorithm in Section \ref{sec:K-means}. We apply these two methods on data generated from Model B in Experiment 3, data generated from Model A in Experiment 4, and data generated from Model C in Experiment 5. The comparison criteria are average error on center values and the average error on all parameters:
\begin{equation}
    \frac{1}{K}\sum_{k=1}^K |\hat{\mu_k} - \mu_k|, \quad  \text{and} \quad
    \frac{1}{K} \sum_{k=1}^K \Big{(} |\hat{\mu_k} - \mu_k| + |\hat{\pi_k} - \pi_k| + |\hat{v_k} - v_k| \Big{)}\ ,
\end{equation}
and the Rand index. Here the sample size is $N=500$, the lower and upper bounds on the separation are $\delta_1=1.9$ and $\delta_2=2.1$ (same across $k$), and the number of repeat is $R=1000$. Note again that the separation constraint is accurate in that it holds in all the models used in this set of experiments.
We record the mean value of all $R$ experiments and their standard deviations in \tabref{Exp_3_4_5}. 
It is clear that in these 3 experiments, the constrained EM algorithm improves over the regular EM algorithm in terms of both parameter estimation and clustering accuracy.

\begin{table}
\centering\small
\caption{Errors and Rand Indices of EM Algorithm and Constrainted EM Algorithm}
\label{tab:Exp_3_4_5}
\bigskip
\setlength{\tabcolsep}{0.15in}
\begin{tabular}{p{0.14\textwidth} p{0.08\textwidth} p{0.2\textwidth} p{0.14\textwidth} p{0.18\textwidth} }
\toprule
%\multicolumn{2}{c}{Item} \\
%\cmidrule(r){1-2}
{\bf Experiment} & {\bf Model} & {\bf Criteria} & {\bf Regular EM} & {\bf Constrained EM} \\ \midrule
Experiment 3 & Model B & center error & 0.339 (0.205) & 0.058 (0.021) \\ 
\text{   } & \text{   } & all parameters error & 0.976 (0.296) & 0.454 (0.197)\\ 
\text{   } & \text{   } & Rand index & 0.893 (0.023) & 0.906 (0.013)\\ \midrule
Experiment 4 & Model A & center error & 0.252 (0.155) & 0.172 (0.120) \\ %\midrule
\text{   } & \text{   } & all parameters error & 0.568 (0.320) & 0.409 (0.242)\\ %\midrule
\text{   } & \text{   } & Rand index & 0.715 (0.047) & 0.726 (0.040)\\ \midrule
Experiment 5 & Model C & center error & 0.448 (0.231) & 0.276 (0.213) \\ %\midrule
\text{   } & \text{   } & all parameters error & 0.994 (0.415) & 0.764 (0.367)\\ %\midrule
\text{   } & \text{   } & Rand index & 0.810 (0.028) & 0.820 (0.030)\\ %\midrule

%Experiment 1 & size error & 165.644 (22.941) & 119.898 (25.439)\\ \midrule
%Experiment 4 & misclassification amount & 87.234 (22.277) & 82.446 (16.526)\\

\bottomrule
\end{tabular}
\end{table}

In the last experiments, we look at the impact of the separation constraint on the result of the constrained algorithms (Experiment 6), and also explore how imposing two-sided constraints differs from imposing one-sided constraints (Experiment 7). 
We focus on the simplest model, Model A, which has only two components. In Experiment 6, we gradually increase $\delta$ on a grid of values ranging from $0.85$ to $2.4$. We generate $R = 100$ data sets from Model A, then for each $\delta$, we fit the model and record all parameters as done in Experiments 3-5. We present the results in $3$ plots with $90\%$ confidence bands in Figure 1.
We can see that the curves representing regular and one-sided EM tend to overlap on the very left. The one-sided EM starts to become more accurate in parameter estimation compared to regular EM when $\delta$ is as low as 0.9, and the improvement increases as we move closer to actual value $2$. Once passed $2$, the performance starts to deteriorate compared to the regular EM. On the other hand, when the two-sided constraint indeed contains the true separation, $2$, the two-sided constrained EM outperforms the one-sided constrained and regular EM. However, quite intuitively, when the two-sided constraint does not include the actual separation inside, the performance of the two-sided constrained EM is significantly worse than the two other algorithms before reaching actual separation value, and the same as the one-sided constrained EM after passing the actual separation value.

\begin{figure}
    \centering
    \centering
    \subfigure[error in the estimation of the centers as a function of the separation $\delta$]{\label{fig:a}\includegraphics[scale=0.33]{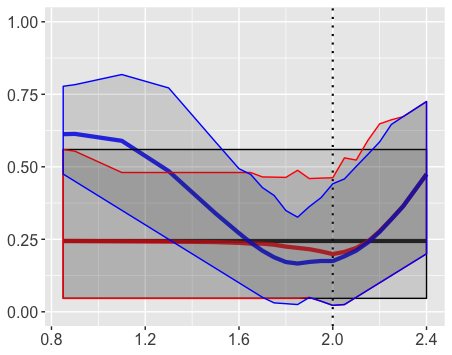}}\hfill
    \centering
    \subfigure[error in the estimation of all the parameters as a function of the separation $\delta$]{\label{fig:b}\includegraphics[scale=0.33]{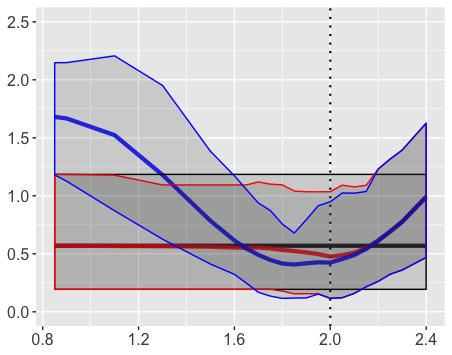}}\hfill
    \centering
    \subfigure[Rand index as a function of the separation $\delta$]{\label{fig:c}\includegraphics[scale=0.33]{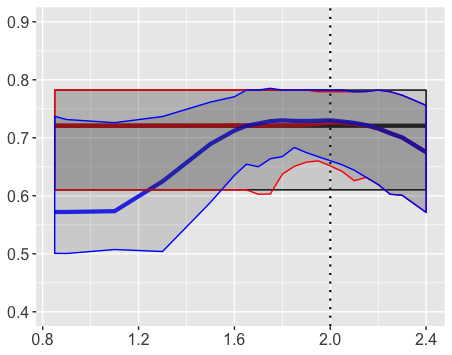}}
    \caption{Results of Experiments 6 and 7: black curves represent unconstrained EM; red curves represent constrained EM with a lower bound on separation, $\delta$; blue curves represent two-sided constrained EM, with lower bound on separation being $\delta_1 = \delta$ and upper bound being $\delta_2 = \delta + 0.2$; thick horizontal line is at the mean values of $100$ experiments, while the thin horizontal lines give the $90\%$ confidence bands; the actual separation, $2$, is indicated by a dotted vertical line. Note that the red and black curves tend to overlap on the left side of plots, and the red and blue curves tend to overlap on the right side of plots.}
    \label{fig:exp67}
\end{figure}

\subsection{Real Dataset}

In this section we apply our constrained EM Algorithm to the 2010 National Youth Physical Activity and Nutrition Study (NYPANS) dataset provided in \citep{CDC2010}. The dataset consists of measurements on high school students in the United States in 2010. Since the height of students in each grade follows an approximate Gaussian distribution, and there are 4 grades, the height of all the students could form a GMM with $K=4$. We focus on estimating the parameters of this GMM.

As prior knowledge, we use information from the data table of stature-for-age charts provided by National Center for Health Statistics in \citep{CDC2001}. The two sources are both gathered in the United States, with a time difference of 9 years. From the mean age of students in the NYPANS study in each of the 4 grades, we found the most relevant median height, and in Gaussian distributions the mean height, in these charts, and subtracted the adjacent values. To account for possible slight changes over the years, we add and subtract respectively $0.005$, in meters, to each of the differences. The resulting prior knowledge, in units of meters, gives us: $\delta_{1,1} = 0.019023, \delta_{2,1} = 0.00895, \delta_{3,1} = 0.00154$ and $\delta_{1,2} = 0.029023, \delta_{2,2} = 0.01895, \delta_{3,2} = 0.01154$. It is clear from \tabref{nypans} that the constrained EM, with proper prior knowledge, outperforms regular EM in both parameter estimation and Rand index, and therefore significantly improves on the ability to uncover the underlying components.

\begin{table}
\centering\small
\caption{Errors and Rand Indices of EM Algorithm and Constrainted EM Algorithm}

\label{tab:nypans}
\bigskip
\setlength{\tabcolsep}{0.15in}
\begin{tabular}{p{0.1\textwidth} p{0.3\textwidth} p{0.15\textwidth} p{0.18\textwidth} }
\toprule
%\multicolumn{2}{c}{Item} \\
%\cmidrule(r){1-2}
{\bf Model} & {\bf Criteria} & {\bf Regular EM} & {\bf Constrained EM} \\ \midrule
NYPANS & center error (times $100$) & 6.7 & 0.5 \\ 
\text{   } & all parameters error (times $100$) & 15.6 & 3.2\\ 
\text{   } & Rand index & 0.566 & 0.596\\

%Experiment 1 & size error & 165.644 (22.941) & 119.898 (25.439)\\ \midrule
%Experiment 4 & misclassification amount & 87.234 (22.277) & 82.446 (16.526)\\

\bottomrule
\end{tabular}
\end{table}

\subsection*{Acknowledgments}
This work was partially supported by a grant from the US National Science Foundation.

\bibliographystyle{chicago}
\bibliography{em.bib}

\appendix

\section{Proof of Proposition 1}
% the \\ insures the section title is centered below the phrase: AppendixA

Using Jensen's inequality, we derive:
\begin{align}
    L(\theta, x) 
    & = \sum_{i=1}^N \log{ \bigg{(}\sum_{k=1}^K \pi_k f(x_i; \mu_k, v_k)\bigg{)} } \\
    & = \sum_{i=1}^N \log{ \bigg{(}\sum_{k=1}^K w_{i,k}^{s+1} \frac{\pi_k f(x_i; \mu_k, v_k)}{w_{i,k}^{s+1}} \bigg{)} } \\
    & \geq \sum_{i=1}^N \sum_{k=1}^K w_{i,k}^{s+1} \log{ \bigg{(}  \frac{\pi_k f(x_i; \mu_k, v_k)}{w_{i,k}^{s+1}} \bigg{)} }  \\
    & = Q(\theta, \theta^s) - \sum_{i=1}^N \sum_{k=1}^K w_{i,k}^{s+1} \log{ (w_{i,k}^{s+1}) } .  
\end{align}
Plugging in $\theta^{s+1}$ for $\theta$, we obtain:
\begin{equation}
    L(\theta^{s+1}, x) \geq Q(\theta^{s+1}, \theta^s) - \sum_{i=1}^N \sum_{k=1}^K w_{i,k}^{s+1} \log{ (w_{i,k}^{s+1}) } ,
\end{equation}
while by definition of the weights in \eqref{E_w}, we have:
\begin{equation}
    L(\theta^{s}, x) = Q(\theta^{s}, \theta^s) - \sum_{i=1}^N \sum_{k=1}^K w_{i,k}^{s+1} \log{ (w_{i,k}^{s+1}) } 
\end{equation}
Thus to prove that $L(\theta^{s+1}, x) \ge L(\theta^s, x)$, it suffices to prove that $Q(\theta^{s+1}, \theta^s) \ge Q(\theta^s, \theta^s)$.
As we maximized $Q(\theta = (\pi, \mu, v), \theta^s)$ in every step of the ECM algorithm with other parameters fixed, we have:
\begin{align}
    Q(\theta^{s+1}, \theta^s) 
    & = Q( (\pi^{s+1}, \mu^{s+1}, {v}^{s+1}), \theta^s ) \\
    & \geq Q( (\pi^{s+1}, \mu^{s+1}, {v}^s), \theta^s) \\
    & \geq Q( (\pi^{s+1}, \mu^s, {v}^s), \theta^s ) \\
    & \geq Q( (\pi^{s}, \mu^s, {v}^s), \theta^s )\\
    & = Q( \theta^s, \theta^s ).
\end{align}

\end{document}